# Active learning-enabled multi-objective design of thermally conductive and mechanically compliant polymers

*Yuhan Liu [1], Jiaxin Xu [2], Renzheng Zhang [2], Meng Jiang [3,4], and Tengfei Luo [2,4,5] ‬*

**AFFILIATIONS**

[1] *Department of Chemistry and Biochemistry, University of Notre Dame, Notre Dame, IN 46556, USA*

[2] *Department of Aerospace and Mechanical Engineering, University of Notre Dame, Notre Dame, IN 46556, USA*

[3] *Department of Computer Science and Engineering, University of Notre Dame, Notre Dame, IN 46556, USA*

[4] *Lucy Family Institute for Data and Society, University of Notre Dame, Notre Dame, IN 46556, USA*

[5] *Department of Chemical and Biomolecular Engineering, University of Notre Dame, Notre Dame, IN 46556, USA*

*: Corresponding authors: tluo@nd.edu

**ABSTRACT**

Polymers are widely used in applications like flexible electronic devices and thermal interface materials due to their mechanical compliance and processing versatility. However, conventional polymers intrinsically have low thermal conductivity (TC), limiting their heat transfer abilities. Moreover, polymer research often focuses on optimizing one specific property, leaving the co-optimization of multiple competing properties less explored. Therefore, identifying polymers that simultaneously achieve high intrinsic TC while maintaining mechanical flexibility (i.e., low modulus) remains a challenge. In this study, we develop an active learning (AL) framework based on multi-objective Bayesian optimization (MOBO) to efficiently discover polymers exhibiting both high TC and low bulk modulus. Initially, a high-throughput molecular dynamics (MD) pipeline was constructed to generate polymer property data, forming a small but informative initial dataset. Independent Deep Kernel Learning (DKL) surrogate models were then constructed for TC and bulk modulus, each integrating a multi-layer perceptron feature extractor with a Gaussian Process model to capture polymer structure-property relationships and quantify prediction uncertainties. These surrogate models guide the $q$NEHVI acquisition function through a larger unlabeled polymer database (~2000 polymers), systematically selecting new polymer candidates for MD simulations and validation. Newly evaluated polymers are iteratively integrated into the training dataset, continuously refining the DKL models to explore the chemical space. Ultimately, six high-performance candidates were identified on the Pareto front, achieving the trade-offs between TC and modulus. Additionally, we applied interpretability techniques to elucidate how molecular structure influences properties under multi-objective trade-offs and assessed the synthesizability of identified candidates. By combining MD simulations with AL-enabled MOBO, our workflow mitigates data scarcity, reduces development time, and provides actionable guidance for designing multifunctional polymers tailored for different applications.

## 1. Introduction

Polymers have become indispensable in industrial applications due to their mechanical flexibility, processing versatility, corrosion resistance, and light weight [1]. Many emerging applications, such as flexible electronics and thermal interface materials, specifically require high thermal conductivity (TC) along with mechanical compliance. However, conventional amorphous polymers exhibit low intrinsic TC, typically below 0.4 $W \cdot m^{-1} \cdot K^{-1}$, due to their highly disordered atomic arrangement [2,3]. This intrinsic limitation significantly constrains their performance in thermal management. Embedding thermally conductive fillers (e.g., ceramics or carbon) into polymer matrices can enhance TC, but the intrinsically low TC of the polymer matrix itself continues to be the primary bottleneck [2,4]. Meanwhile, flexible applications demand polymers with mechanical compliance (i.e., low bulk modulus) to facilitate conformal contact and mitigate mechanical stresses in flexible assemblies. In thermal interface materials, the mechanical compliance is also critical to fill the microscale gaps between two flat surfaces. Therefore, there is a compelling need to identify polymers that simultaneously possess high intrinsic TC while maintaining softness.

Traditionally, the discovery and design of new polymers have relied heavily on intuition-driven experimentation and trial-and-error discovery. However, inconsistent synthesis processes and measurement methods often result in prolonged testing periods and significant costs, limiting the efficiency and scalability of polymer screening. For example, PoLyInfo[5], one of the largest publicly accessible polymer databases, provides sparse experimentally measured data for neat homopolymers, with only 91 polymers for TC and 36 for bulk modulus. For some polymers, significant variations persist even under comparable measurement temperatures. For instance, the reported TC for polypropylene ranges from 0.175 to 0.78 W·m$^{-1}$·K$^{-1}$, while the bulk modulus of polystyrene spans 3.55 to 11.1 GPa. Such substantial inter-source variability introduces large noise into data-driven modeling and analysis. Molecular dynamics (MD) simulations provide valuable computational insight into polymer structure-property relationships but are limited by their high computational cost when applied at scale[6,7]. Recent advancements in polymer informatics (PI) offer a promising route toward accelerating polymer discovery through data-driven methods[8,9,10,11]. However, typical machine learning (ML) pipelines depend heavily on large, high-fidelity datasets[12] and typically focus on optimizing single properties individually[13,14]. In contrast, practical polymer applications frequently necessitate the simultaneous optimization of multiple, sometimes competing attributes. Specifically, increasing intrinsic TC typically involves rigid, ordered, highly aligned polymer backbones to enhance phonon transport[15,16,17], thereby increasing stiffness. Conversely, achieving a low modulus often requires flexible, less intermolecularly constrained molecular chains[18,19,20], which can impede thermal transport. Thus, enhancing one property often comes at the expense of another, which requires explicitly balancing these intrinsic trade-offs through targeted multi-objective optimization strategies in polymer design.

Recently, significant research efforts have focused on employing active learning (AL) for data-efficient, single-objective polymer discovery[21,22,23], as well as for multi-objective optimization across broader classes of materials[24,25,26,27]. AL, a ML approach that strategically selects the most informative data points from an unlabeled pool for labeling[28], is particularly attractive when high-fidelity property evaluation is computationally or experimentally expensive. However, AL frameworks that explicitly target multifunctional polymer discovery remain scarce[29,30]. Among available data-driven optimization strategies, Bayesian optimization (BO), typically built on Gaussian process (GP)-based surrogate models, can efficiently guide the selection of promising candidates by balancing exploration and exploitation via designed acquisition functions[22,23,31]. Extending BO into multi-objective Bayesian optimization (MOBO) enables the identification of optimal trade-offs among conflicting properties, generating a set of non-dominated solutions known as the Pareto front (i.e., solutions where one property cannot be improved without compromising the other). Integrating AL with MOBO can address data scarcity issues and further accelerate polymer screening by prioritizing candidates expected to maximally reduce model uncertainty, thereby improving surrogate accuracy with minimal additional high-fidelity data, which is particularly valuable for multifunctional polymer design.

In this study, we develop an AL-enabled MOBO workflow for identifying amorphous polymers with simultaneously high TC ($k$) and low bulk modulus ($B$), as illustrated in Fig. 1. First, a high-throughput MD pipeline is established to generate polymer property labels ($k$ and $B$), constructing a small initial dataset. Next, separate Deep Kernel Learning (DKL) surrogate models for $k$ and $B$ are constructed to effectively capture polymer structure-property relationships, each consisting of a multi-layer perceptron (MLP) feature extractor coupled to a GP model for uncertainty-informed prediction. These surrogates then screen a large unlabeled polymer pool. Guided by predictive means and quantified uncertainties, an acquisition function selects polymer candidates expected to yield the greatest improvements in the Pareto front. The selected polymers are evaluated through MD simulations, and the resulting new data are integrated to continuously refine the surrogate models and inform the next selection, thus forming a closed optimization loop. By repeating this AL-MOBO loop, we systematically advance the Pareto front and ultimately identify six high-performance candidates with optimal high-$k$ and low-$B$ characteristics. Beyond performance evaluation, we integrate physical descriptors with interpretable analysis to uncover molecular-level structure-property relationships governing multi-objective trade-offs and further assess synthesizability of identified polymers. By integrating MD simulations, AL-enhanced MOBO, and uncertainty-aware surrogate models, our workflow offers a data-efficient pathway that provides guidance for the development of multifunctional polymers for target applications.

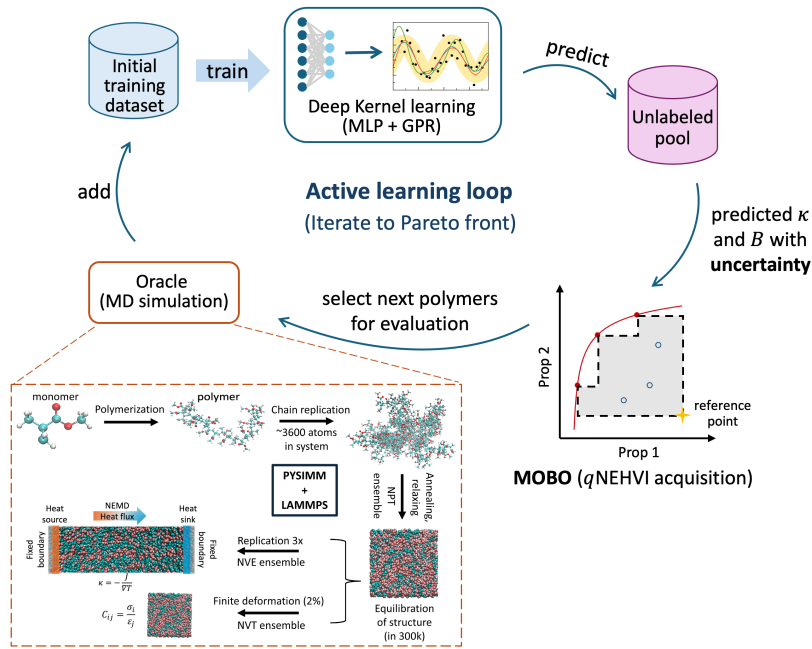

**Figure 1.** Schematic of the active learning multi-objective Bayesian optimization (AL-MOBO) workflow to discover amorphous polymers with high thermal conductivity (TC, $k$) and low bulk modulus ($B$). A high-throughput MD "oracle" (bottom left) provides $k$ and $B$ labels for an initial training set and for iteratively selected candidates; uncertainty-aware deep kernel learning (DKL) surrogates (top) map polymer structure to $k$ and $B$, and a $q$-Noisy Expected Hypervolume Improvement ($q$NEHVI)-based MOBO process (right) uses their predictions to choose new polymers from the unlabeled pool and advance the $k$-$B$ Pareto front.

## 2. Results and discussions
### 2.1 Dataset

We curated an unlabeled screening pool of over 2000 homopolymer structures, sourced from the publicly accessible PoLyInfo[5] and Membrane Society of Australasia (MSA)[32] Polymer Membrane databases. This polymer dataset exhibits chemical diversity, covering a wide range of backbone families, including polyesters, polyamides, polyolefins, polyethers, polyketones, and other major classes represented in PoLyInfo[6]. Polymer structures are encoded using polymer SMILES (p-SMILES)[9, 33], capturing monomer composition and polymerization connectivity, subsequently converted into numerical polymer embeddings (PE)[9] as model inputs. From the unlabeled set, we selected an initial subset of 93 polymers as training data for surrogate modeling. To ensure representative coverage and avoid bias toward specific chemistries, we utilized a Latin Hypercube Sampling (LHS) strategy with a maximin criterion for optimal space-filling[34]. TC and bulk modulus values for these polymers were computed through MD simulations, which is detailed in the Method Section 4.1. As illustrated in Fig. 2a and 2b, the initial dataset spans a TC range of 0.047 - 0.551 W·m$^{-1}$·K$^{-1}$ and a bulk modulus range of 1.823 - 6.989 GPa. The high-dimensional polymer embedding space was qualitatively visualized using t-distributed Stochastic Neighbor Embedding (t-SNE)[35] (Fig. 2c). In addition, the polymer-family distribution of the initial set relative to the unlabeled pool is summarized in Supplementary Table S1 and Fig. S1. These results indicate that the initial training set exhibits broad coverage across the chemical space, avoiding early bias toward specific chemistry while leaving sparsely populated regions for adaptive exploration by the acquisition function during MOBO.

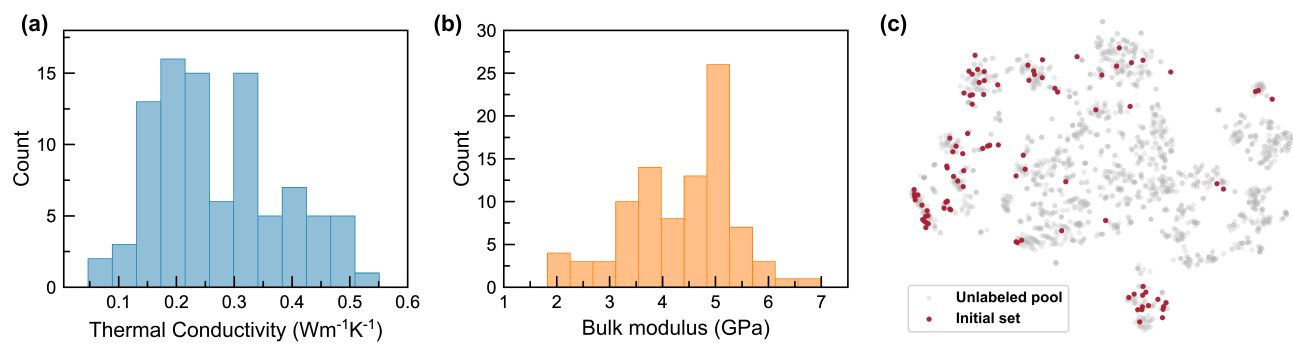

**Figure 2.** Overview of the initial dataset. (a) TC and (b) bulk modulus distributions for the 93 amorphous polymers in the initial MD-labeled training set. (c) t-SNE visualization of the polymer embedding space, showing the initial set (red) relative to the unlabeled screening pool (gray).

## 2.2 Validation of polymer MD simulations

### 2.2.1 Comparison with experimental data

In this work, MD simulations were employed as the data-generation oracle within the AL framework to ensure all property labels were produced under consistent procedures. To validate the reliability of the MD simulations, polymers with experimentally reported property values were collected from the PoLyInfo database and other publicly available literature[36,37]. For fair comparison, only measurements obtained between 20 °C and 30 °C were considered, corresponding to room-temperature simulation conditions. The validation set comprised 28 TC and 18 bulk modulus data points. As shown in Fig. 3a and 3b, the MD-predicted values exhibit good agreement with experimental measurements, yielding coefficients of determination ($R^2 = 0.751$ for TC and $R^2 = 0.641$ for bulk modulus). The observed deviations likely arise from variations in experimental conditions and measurement protocols, as well as residual uncertainty in the MD simulations due to finite sampling, which together have been found to introduce substantial scatter in MD-experiment comparisons[38].

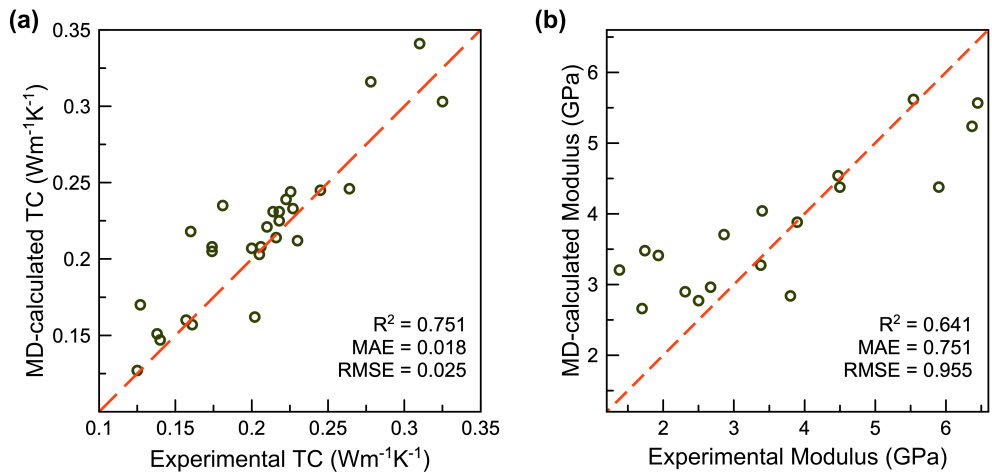

**Figure 3.** Experimental validation of the MD oracle. Parity plots comparing MD-calculated values against experimental measurements for (a) TC and (b) bulk modulus.

### 2.2.2 Influence of polymer morphology

Previous studies have verified that randomness in polymer morphology during structure generation can introduce variations of 7-20% in the simulated TC of amorphous polymers[38]. To assess whether similar effects impact modulus evaluation, we examined the influence of initial structural configurations on the calculated bulk modulus. Three independent initial structures were generated for each of seven common polymers, and the resulting moduli were compared. As shown in Supplementary Fig. S2a, the coefficients of variation (defined as the ratio of standard deviation to the mean) within a single simulation were all ≤ 0.334, whereas the values across simulations with different initial structures remained below 0.041. These small deviations indicate that modulus calculations are largely insensitive to morphological randomness in the initial polymer configurations. However, to eliminate potential morphology-related biases for both TC and bulk modulus, all MD-calculated property values reported in this study were averaged from simulations conducted on three distinct initial structures for each polymer.

### 2.3 The AL-MOBO framework for multifunctional polymers
#### 2.3.1 Polymer representation comparison

Prior to training the surrogate model, it is essential to establish a numerical representation for polymers to accurately quantify structure-property relationships. To identify the most informative representation, we benchmarked four commonly used methods: PE, Morgan Fingerprints (MF)[39], MACCS Keys[40], and RDKit Fingerprints[41], on the initial dataset of 93 polymers using a unified 5-fold cross-validation (CV). Comparison was conducted under a fixed GP regressor, consistent with our subsequent DKL setup, as detailed in the Supplementary Information (SI). A GPR baseline was selected over DKL to attribute performance differences solely to the representation, avoiding confounding factors introduced by MLP, such as additional hyperparameters, neural architectures, and stochastic optimization, which may obscure the true representation effects, particularly on small datasets. As summarized in Supplementary Table S2, PE achieved the highest predictive accuracy for both TC ($R^2 = 0.757$, mean squared error, MSE = 0.003) and bulk modulus ($R^2 = 0.551$, MSE = 0.459). PE, inspired by the word2vec concept[9,42], represents p-SMILES into continuous-valued 300-dimensional vectors. This representation effectively captures subtle chemical differences influencing polymer properties, enabling more reliable property predictions[43]. Consequently, PE was adopted as the polymer representation for subsequent DKL surrogate modeling.

### 2.3.2 Construction of surrogate models

As PE are high-dimensional embeddings, Principal Component Analysis (PCA) was applied to reduce dimensionality before input into Gaussian Process (GP) regression, mitigating the curse of dimensionality[23]. However, as an unsupervised method, PCA discards response-relevant variance due to the lack of target-informed projections, adversely affecting predictions reliant on nonlinear features. While PCA was found effective for TC in our previous work[23], the present study observed that PCA preprocessing significantly diminished critical information necessary for accurately predicting bulk modulus. Specifically, the 5-fold CV $R^2$ for the GP model using PCA declined from approximately 0.6 initially to nearly 0.1 in subsequent iterations (Supplementary Fig. S3a), indicating severe degradation in predictive accuracy and generalization as optimization proceeded. Such feature loss decreases the signal-to-noise ratio, impairing the accurate capture of structure-modulus relationships, likely due to the strong dependence of bulk modulus on interchain interactions.

To address this, we employed DKL surrogate models, trained independently for TC ($k$) and bulk modulus ($B$), incorporating MLP encoders prior to GP regression. These MLP encoders compressed the original 300-dimensional PE into compact latent spaces (16-dimensional for $k$ and 12-dimensional for $B$), effectively preserving essential nonlinear information (hyperparameter optimization details provided in the Method Section 4.2). This approach effectively circumvented both the curse of dimensionality and loss of predictive features. The latent representations were subsequently input to GP regressors, providing predictive means and uncertainties. As illustrated in Supplementary Fig. S3b, DKL models demonstrated consistently robust performance throughout the AL campaign. Specifically, TC surrogate maintained stable predictive accuracy with $R^2$ ~0.8 and consistently low MSE, while the bulk modulus surrogate displayed steady performance with CV $R^2$ around 0.58, without any noticeable deterioration. The initial surrogate accuracy, presented in Fig. 4, exhibited good agreement in trend between predictions and MD ground truth labels: TC predictions achieved CV $R^2 = 0.804$ and MSE = 0.002 (Fig. 4a), and bulk modulus predictions yielded CV $R^2 = 0.659$ and MSE = 0.348 (Fig. 4c). These results demonstrate the efficacy of DKL surrogates in capturing the structure-property relations of both target properties even with limited initial data.

Subsequently, these DKL models were utilized to predict properties for the entire unlabeled polymer pool. Fig. 4b and 4d visualize predicted means with posterior uncertainties ($\sigma$), represented as vertical bars. Moreover, initial uncertainty was quantified using 5-fold out-of-fold predictions via Spearman rank correlation and empirical coverage to evaluate uncertainty discrimination and calibration, as reported in Supplementary Table S3. These diagnostics motivate the subsequent assessment of uncertainty calibration. These posterior estimates guide the exploration-exploitation strategies during candidate selection, where high predictive means drive exploitation of promising candidates, and large posterior variances indicate under-characterized regions for exploration. The $q$-Noisy Expected Hypervolume Improvement ($q$NEHVI) acquisition function balances exploration and exploitation by maximizing the expected hypervolume improvement under the joint posterior, yielding the selection of informative candidates for subsequent iterations.

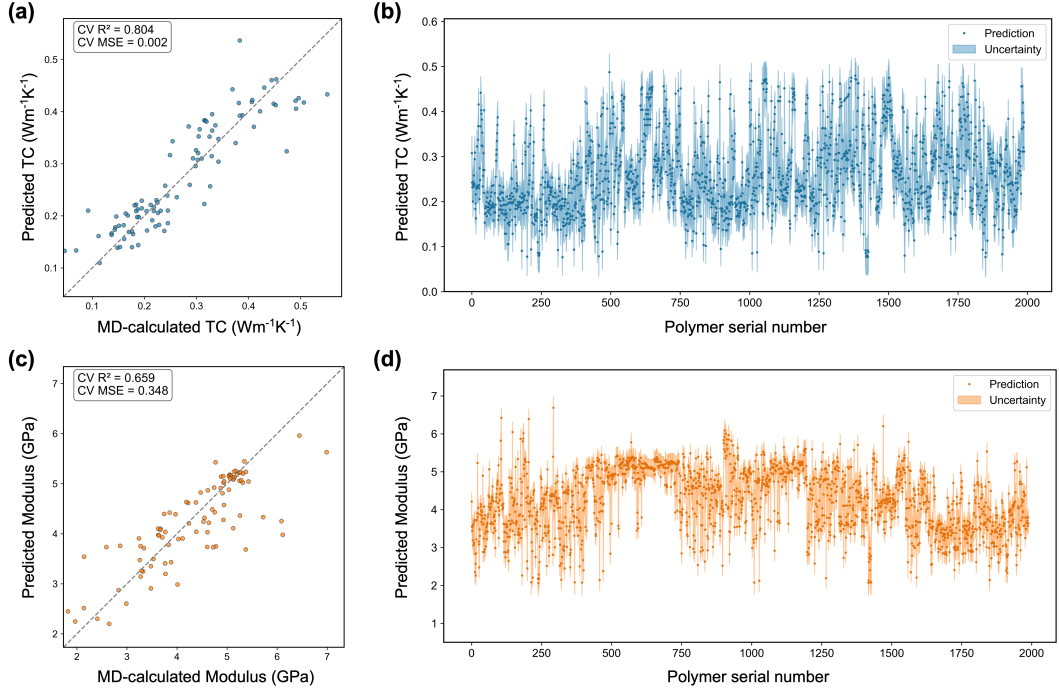

**Figure 4.** Performance of the DKL surrogates and their initial predictions at iteration 0. 5-fold cross-validated parity plots comparing DKL-predicted and MD-labeled values for (a) TC and (c) bulk modulus. (b) and (d) show the DKL-predicted means (points) with predictive uncertainties (vertical bars) over the unlabeled polymer pool for TC and bulk modulus, respectively.

### 2.3.3 Performance assessment of the AL-MOBO framework

To effectively balance exploration and exploitation under noise during optimization loop, we employed the $q$NEHVI algorithm [44] as our acquisition function. At each iteration, we concurrently selected four polymer candidates ($q = 4$) to maximize the joint expected hypervolume gain while accounting for observation noise (see Method Section 4.3 for details). This promotes diverse and complementary candidate selections, optimizing guidance while managing computational costs. Throughout the 60 iterations performed (240 evaluated polymers), the AL-MOBO performance was quantified using hypervolume (HV) and incremental improvements per iteration (ΔHV). Additionally, surrogate reliability and uncertainty calibration were monitored via negative log likelihood (NLL) and expected normalized calibration error (ENCE). We detail below the Pareto front development, optimization trajectory, sample efficiency via HV metrics, as well as model reliability and calibration.

Fig. 5a-c summarizes the optimization progress, showing rapid early improvements followed by diminishing gains and practical convergence around iteration 31. To verify convergence within the candidate pool, we continued optimization until iteration 60. Fig. 5a illustrates the evolution of the Pareto set through three representative snapshots, highlighting an expand-then-contract pattern: the number of non-dominated polymers peaked at 11 by iteration 8, then contracted to 6 by iteration 31 as superior candidates replaced earlier suboptimal ones. Fig. 5b depicts the sampling trajectory across all 60 iterations, chronologically color-coded, illustrating a shift from the initial broad exploration, driven by higher uncertainty, to targeted exploitation near the evolving Pareto front. Because the candidate pool is discrete, the trade-off region is not uniformly populated. Later iterations (lighter points in Fig. 5b) further sampled near the Pareto front, particularly the mid-range region, to densify local coverage and confirming convergence. The final Pareto set (red stars) consists of six polymers identified at iterations 6, 13, 19, 29, and 31, covering TC from 0.127 to 0.637 W m$^{-1}$ K$^{-1}$ and bulk modulus from 1.09 to 4.654 GPa. These selections effectively represent diverse and complementary trade-offs, identifying the global extremes (lowest modulus and highest TC) among all MD-labeled polymers.

Fig. 5c evaluates sample efficiency through HV and its per-iteration increment (ΔHV). HV measures the volume of space dominated by the current Pareto front relative to a fixed reference point (in two objectives, this corresponds to the dominated area)[45]. The HV progression reveals three distinct phases: rapid early gains (iterations 1-6), a subsequent plateau with minimal improvements (iterations 7-28), and a notable surge around iterations 29-31 followed by stabilization. HV increases only when newly evaluated polymers expand the non-dominated region (discrete, stepwise jumps rather than smooth changes). This behavior aligns with $q$NEHVI, which jointly selects each batch to maximize the expected HV gain. Consequently, ΔHV is episodic rather than continuous. Sparse peaks in ΔHV indicate iterations where the Pareto front shifts outward, producing the observed HV improvements; moreover, it serves as a practical stopping criterion, signaling convergence around iteration 31. Early HV gains primarily arise from identifying polymers with notably low moduli, while later enhancements result from discovering polymers with the highest TC and strategically filling intermediate gaps. The best-so-far TC increased substantially from 0.482 W m$^{-1}$ K$^{-1}$ at iteration 1 to 0.637 W m$^{-1}$ K$^{-1}$ by iteration 29 (a 32% improvement), while the best bulk modulus decreased notably from 1.892 GPa to 1.09 GPa by iteration 6 (a 42% reduction). Both metrics subsequently remained stable (Supplementary Fig. S4a), consistent with the observed convergence of the Pareto front.

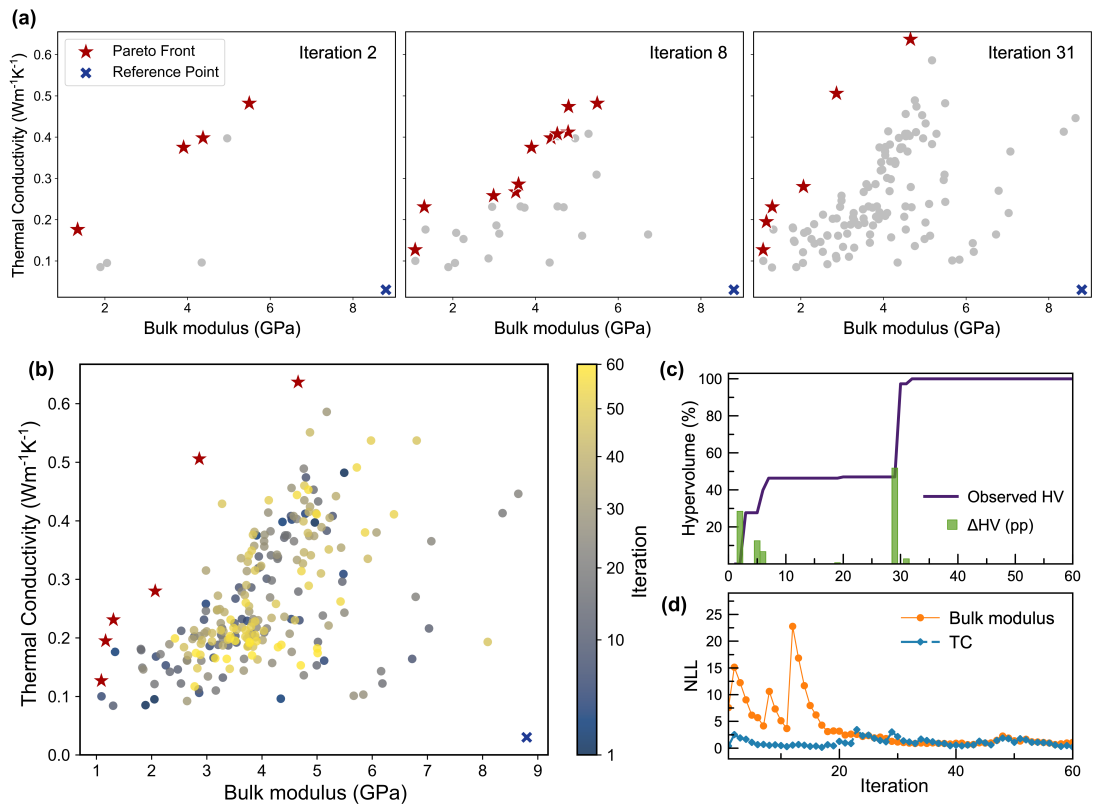

**Figure 5.** Optimization trajectory and surrogate reliability of the AL-MOBO strategy. (a) Snapshots of the non-dominated (Pareto) set at iterations 2, 8, and 31 (red stars) overlaid on all polymers evaluated up to that iteration (gray). The fixed reference point (blue cross) is set to be dominated by all evaluated polymers. (b) Sampling trajectory over 60 iterations, with AL-selected polymers color-coded by iteration index (1-60); the final Pareto set is marked by red stars. (c) Observed hypervolume rescaled to 0-100% (line) and the corresponding per-iteration increment (ΔHV) reported in percentage-point (pp) (bars). (d) EWMA-smoothed negative log-likelihood (NLL) for $k$ and $B$ surrogates across iterations.

Surrogate model reliability was primarily assessed using the negative log likelihood (NLL), which penalizes both inaccurate predictions and miscalibrations (over- and under-confidence). As displayed in Fig. 5d, NLL decreased for both surrogates and plateaued around iteration 30, coinciding with the observed HV and Pareto front stabilization. Across 60 iterations, the bulk modulus surrogate initially exhibited high NLL values due to its inherently noisier and more challenging prediction compared to TC. Nevertheless, its NLL sharply decreased from an initial average of 9.51 (iterations 1-5) to 1.03 over the final ten iterations, representing an approximately 89% reduction. The TC surrogate exhibited lower and less variable NLL, decreasing from 1.77 initially to 0.74 over the same interval (58% reduction) and reaching 0.21 by the end of the campaign. Uncertainty calibration was further assessed using the expected normalized calibration error (ENCE), which measures the alignment between predicted uncertainties and actual errors. As shown in Supplementary Fig. S4b, rolling-pooled ENCE (5-round window) decreased from 1.05 to 0.63 for TC (40% reduction) and from 3.29 to 0.24 for bulk modulus (93% reduction), highlighting progressive calibration improvements.

Collectively, these results indicate enhanced posterior predictive distributions characterized by improved predictive accuracy (lower NLL) and better-calibrated uncertainties (lower ENCE). This continuous surrogate refinement underpinned the transition from initial exploration to reliable convergence. Together with steadily increasing HV, these findings validate the effectiveness of the AL-MOBO framework in efficiently exploring the polymer design space.

## 2.4 Post-optimization interpretability and analyses
### 2.4.1 t-SNE visualization of polymer space

To qualitatively show how AL-MOBO expanded coverage and converged near Pareto-relevant regions, we visualized the chemical embedding space using t-SNE (Fig. 6a). Specifically, we projected the 300-dimensional PE into two dimensions using t-SNE, overlaying four cohorts: the unlabeled polymer pool (gray), the initial dataset (red), all MOBO-selected candidates (color-coded by iteration from 1 to 60), and the final Pareto candidates (blue stars). The resulting t-SNE map shows a progressive optimization trajectory. Early iterations primarily sample neighborhoods close to the initial dataset, reflecting cautious exploration around known structures. As iterations advance, candidates gradually extend into previously under-sampled regions of the chemical embedding space, indicative of broader exploratory behavior. In later iterations, the sampled polymers increasingly concentrate in several distinct neighborhoods that coincide with the Pareto-optimal solutions. Notably, these Pareto solutions are not confined to a single region but instead span multiple disconnected clusters, some of which were not covered by the initial set. This distribution highlights the capability of MOBO to uncover diverse and complementary polymer chemotypes. Moreover, many late-iteration selections appear near the boundaries of densely populated regions of the unlabeled pool, implying strategic exploitation at the frontiers rather than solely within densely sampled areas.

Since t-SNE preserves local neighborhood relationships without faithfully representing global cluster sizes or inter-cluster distances, we interpret the spatial patterns qualitatively: the proximity of late-iteration points to Pareto candidates signifies neighborhood-level enrichment, whereas absolute distances are not over-interpreted. Thus, the t-SNE visualization complements quantitative assessments (such as hypervolume trajectory) by qualitatively demonstrating where the transition from exploration to exploitation occurs. Overall, the visualization confirms that MOBO expanded beyond initial regions, identified multiple structurally diverse high-value clusters, and effectively exploited previously under-explored areas, thereby providing spatial context consistent with observed improvements in optimization performance.

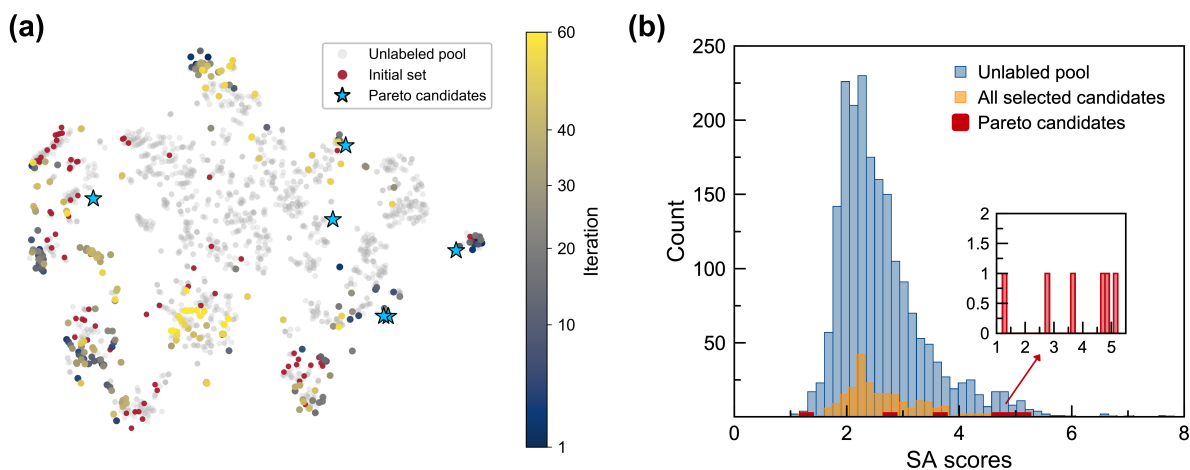

**Figure 6.** Exploration of polymer chemical space and synthesizability assessment. (a) t-SNE projection of the polymer embedding space showing the unlabeled pool (gray), initial set (red), all MOBO-selected candidates (colored by iteration), with Pareto polymers highlighted as blue stars. (b) Distributions of SA score for the unlabeled pool (blue), all selected candidates (orange), and Pareto polymers (red), where all three are concentrated in the easy-to-moderate range.

### 2.4.2 Synthesizability evaluation

Schuffenhauer's synthetic accessibility (SA) score[46] is widely used to quantify molecular synthesizability on a scale from 1 (easy) to 10 (difficult), with scores above 6 generally considered challenging. Here, we analyzed SA score distributions for three datasets: the unlabeled pool, all MOBO-selected candidates, and the final Pareto set (Fig. 6b). Both the unlabeled pool and all selected candidates exhibit unimodal distributions centered around the easy-to-moderate region (SA ≈ 2-4), with very few high-difficulty cases (SA > 6). Compared to the unlabeled pool, the selected candidates show a slight shift toward higher complexity, reflected by a modest increase in median (2.42 to 2.49) and mean (2.61 to 2.79). However, the proportion of difficult-to-synthesize polymers remains low ($P_{SA > 6}$ = 0.021). Moreover, while the final Pareto candidates exhibited higher structural complexity (median 4.16), all remained within a reasonable synthesis range, with SA scores spanning 1.21 to 5.13. These results indicate that the MOBO framework effectively explored a more complex chemical space without exceeding the threshold of synthetic accessibility, supporting the practical feasibility of the identified Pareto optimal polymers for future experimental validation.

### 2.4.3 Structure-property interpretability analysis

In our MOBO workflow, opaque embeddings inherently restrict direct insight into how specific structural features influence polymer properties. To obtain mechanistic insights beyond the "black-box" nature of candidate selections[47], we utilized physically interpretable descriptors to train explainable tree-based models (performance details provided in the SI), thereby explicitly attributing property variance to specific structural features. Specifically, we employed hierarchical descriptors from PolyMetriX[48], minimally complemented by first-order RDKit physicochemical descriptors, which distinguish contributions from polymer backbone and sidechain attributes. These descriptors explicitly represent chemical and topological characteristics, such as backbone rigidity and π-conjugation, chain flexibility and saturation, polarity and hydrogen bonding, charge, and proxies for packing or polarizability[49].

To establish interpretable structure-property relationships, we quantitatively mapped descriptors to property contributions using SHapley Additive exPlanations (SHAP)[50]. For improved readability, SHAP values are reported using a consistent sign convention (i.e., positive SHAP values indicate an increase in the predicted TC or bulk modulus). Furthermore, SHAP plots consistently differentiate Backbone (B) and Sidechain (S) attributes (Fig. 7), directly linking structural hierarchies to property findings.

Global beeswarm plots (Fig. 7a and 7b), derived from all MD-labeled polymers (306 polymers), highlight the most influential B/S descriptors and their directional influences on polymer properties (descriptor definitions are provided in the Supplementary Table S4). Overall, backbone rigidity and π-conjugation metrics (e.g., AromaticFrac, AroRings, Balaban J) contribute positively to the SHAP values for both TC and bulk modulus. This finding agrees with previous studies showing that rigid, ordered backbones enhance phonon transport, especially along the backbone[17,51], thereby improving TC at the cost of mechanical flexibility. Conversely, the descriptor FracCSP3 (fraction of sp³ carbons) represents saturated carbon content. Within our amorphous polymer dataset, higher FracCSP3 (B) corresponds to increased rotatable dihedrals, causing segmental disorder and phonon scattering, which reduces TC, consistent with MD observations that chain rotation lowers the TC of single-chain polymers[52,53]. Additionally, sidechain flexibility (captured by higher FracCSP3 (S)) decreases the modulus by enhancing segmental mobility and loosening packing, aligning with studies indicating longer, more flexible alkyl side chains soften polymers[54,55]. Collectively, these insights confirm the anticipated trade-off between polymer rigidity and mechanical compliance.

Polarity-related descriptors further differentiate the mechanisms governing bulk modulus and TC. TPSA (B/S), representing polar surface area[56], exhibits positive contributions to bulk modulus, consistent with stronger noncovalent interactions (e.g., dipolar and hydrogen-bonding) enhancing interchain electrostatic cohesion and thereby increasing modulus[57]. For TC, although thermal transport is often reported to improve when intermolecular coupling (e.g., H-bonds) is strengthened in polymers[58], our amorphous dataset shows a negative net effect of polarity: depending on chain conformation and packing state, electrostatic interactions among polar groups can promote self-association and more compact conformations[59] (e.g., smaller radius of gyration), which may impede effective backbone-mediated thermal transport, leading to lower TC[17]. This interpretation aligns with the observed negative SHAP contribution of MaxAbsQ (S) to TC. Notably, nO (B) acts as a composite proxy: beyond reflecting polymer polarity, it also correlates with oxygen-containing linkage types and associated conformational flexibility (e.g., softer torsional potentials), which likewise contributes negatively to TC in this dataset. Overall, lowering polarity density emerges as a feasible lever for increasing TC while decreasing bulk modulus.

Additional descriptors for TC and bulk modulus further refine these mechanistic interpretations. Regarding TC (Fig. 7a), sidechain complexity indices, such as SidechainDiv (S) and NumSidechains (S), negatively correlate with TC. This is likely related to the vibrational interaction between the side chain and the backbone: the diverse sidechains can have diverse vibrational frequency spectra, which scatter different phonon modes transporting along the backbone, thus reducing TC[60]. Additionally, high HalogenFrac (S) aligns with lower TC, consistent with increased mass-disorder and phonon scattering as well as reduced phonon group velocities in heavy-atom-containing polymers[61]. For modulus descriptors (Fig. 7b), features that improve packability tend to increase bulk modulus (reducing polymer flexibility), whereas packing-disruptive features decrease bulk modulus and thus promote flexibility. Specifically, Kappa2 (S) reveals that more rod-like, less-branched side chains pack more efficiently, which can elevate modulus[54]. Most visibly, LabuteASA (B), a 2D proxy for backbone accessible surface, illustrates that larger contactable backbone area is associated with a higher modulus, consistent with the classical view that greater inter-chain packing corresponds to higher stiffness [62]. Conversely, the descriptor AliphRings (B) reflects a softening trend in this dataset, consistent with cycloaliphatic (contorted) units disrupting backbone planarity and inter-chain packing[63,64].

Overall, SHAP analysis indicates that TC is governed primarily by intrachain vibrational transport along the covalently connected backbone, favored by straighter and more rigid backbones, whereas the bulk modulus is dominated by interchain cohesion and packing, reflected by polarity and accessible-surface descriptors. These insights offer actionable guidance for multi-objective polymer design: maintaining rigid, π-rich backbones to enhance TC, while reducing modulus primarily through non-polar flexibility (e.g., rotatable segments at side chains) and backbone contortions that impede dense packing. Simultaneously, strongly polar side chains should be avoided, as these negatively impact TC in our models.

Within these structure-property trends, the six polymer candidates ultimately identified on the Pareto front (Fig. 7c) exemplify distinct structural strategies to balance TC and bulk modulus. PF5 (polyethylene) anchors the high-TC boundary: its chemically simple, fully saturated hydrocarbon backbone facilitates efficient packing, yielding the highest TC in our dataset (0.637 W m$^{-1}$ K$^{-1}$), though at the expense of a high bulk modulus, highlighting the stiffness penalty at this extreme. Conversely, PF1 [poly(1,2,2-trifluorovinyl 1,1,2,2,3,4,4-heptafluorobut-3-enyl ether)] occupies the opposite end of the Pareto front with extensive fluorination and a small ether linkage. The strong C-F bonds combined with low polarizability lead to low interaction energy between chains and thus the lowest modulus observed (1.09 GPa).

Positioned between these extremes, PF3 [poly(3,3′-bi(11H-5-thia-6,12-diazananaphthacene)-8,8′-diyl] features a π-conjugated backbone containing N/S heteroatoms. Its rigid π-framework leads to relatively high TC, while potential hydrogen-bonding N–H/amine sites tend to enhance interchain cohesion, placing its modulus towards the higher end of the Pareto front. PF2 (PIM-PI-1)[65] and PF4 (PIM-PI-4)[65], recognized as polyimides of intrinsic microporosity (PIM-PIs), incorporate locally rigid π-rich segments combined with spiro-center-induced geometric contortion. Consistent with classical PIM topologies, these "rigid yet contorted" structures inhibit dense packing and produce high free volume[66,67]. Additionally, many of the O/N sites appear sterically shielded within their bulky fused frameworks, limiting the formation of extended inter-chain hydrogen-bonding networks, reflected in our simulations as relatively low moduli paired with moderate TC values. Lastly, PF6 (PBO-PI 0/100) is a rigid, aromatic polymer belonging to the PBO-PI family, characterized by multiple aryl-O-aryl ether linkages. Its π-rich but kinked fused aromatic backbone packs less efficiently than PF5, resulting in intermediate modulus values between the softer PIM-PIs and the tightly packed hydrocarbon (PF5). Collectively, these examples support our established design principles, demonstrating structurally diverse, chemically complementary polymers capable of achieving a range of property trade-offs.

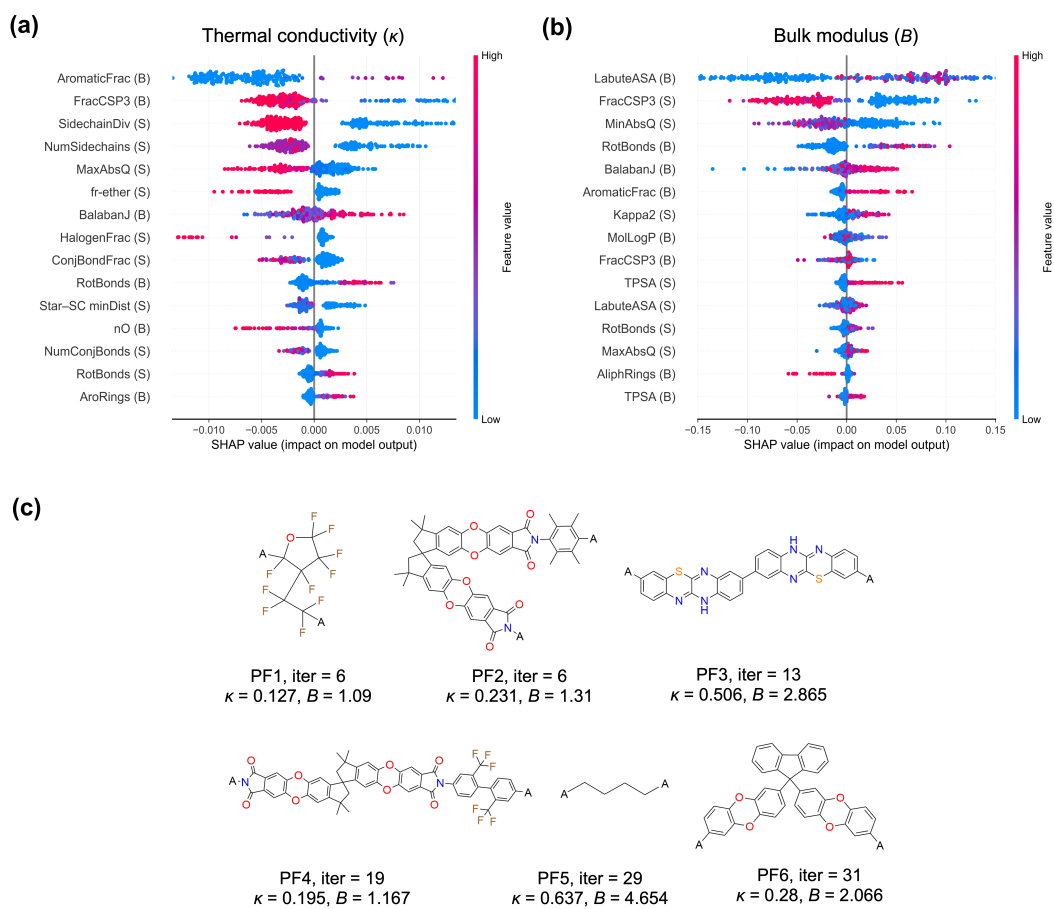

**Figure 7.** SHAP-based interpretability and structures of Pareto-optimal polymers. (a, b) SHAP beeswarm plots ranking the top-15 backbone (B) and sidechain (S) descriptors governing (a) TC and (b) bulk modulus. Positive values indicate an increase in the target property. Points are colored by standardized feature value (red = high, blue = low). (c) Chemical structures and MD-predicted TC (W·m$^{-1}$·K$^{-1}$) and bulk modulus

(GPa) of the six Pareto-optimal polymers identified by AL-MOBO, illustrating diverse motifs governing the optimal trade-offs.

## 3. Conclusion

This study presents an AL-enabled MOBO framework designed to discover amorphous polymers that simultaneously achieve high TC and low bulk modulus. By integrating a high-throughput molecular dynamics pipeline, deep-kernel learning surrogates, and the $q$NEHVI acquisition strategy, the framework efficiently explored the polymer space and identified six non-dominated candidates, which span the performance extremes (lowest modulus and highest TC) and capture the trade-off between mechanical compliance and heat transport efficiency. Structure-property interpretability analyses further elucidated the molecular mechanisms underlying this trade-off, revealing how intrachain backbone rigidity and interchain cohesive interactions jointly influence the balance between thermal transport and elastic stiffness. The synthetic accessibility assessment indicated that the identified Pareto polymers are both chemically diverse and practically attainable. Looking forward, this scalable workflow offers a pathway to explore additional properties and material classes. Future efforts will involve applying this framework to larger polymer libraries (e.g., entire PoLyInfo, PI1M databases) and enhancing molecular dynamics accuracy to further expand discovery potential and predictive reliability. Overall, this study establishes a data-efficient and interpretable paradigm for the systematic design of multifunctional polymers, offering guidance for targeted materials design in applications such as flexible electronics and thermal interface materials.

## 4. Methods
### 4.1 MD simulation

TC ($\kappa$) and bulk modulus ($B$) were obtained via a two-step MD simulation comprising amorphous structure generation and optimization, followed by property evaluation, which are Non-Equilibrium MD (NEMD) for $\kappa$ and finite-strain elasticity for $B$.

**Amorphous Polymer Structure Generation and Optimization.** Our methodology involves the polymerization of monomers, chain replication, and fully structural relaxation. Polymer monomers are represented in SMILES (Simplified Molecular Input Line Entry System) strings and serve as inputs to PYSIMM[68] (a Python-based automated pipeline). As shown in Fig.

1, we use this pipeline to automate the polymerization of monomers into polymer chains with approximately 600 atoms each[69]. The General AMBER Force Field 2 (GAFF2)[70] forcefield parameters are assigned simultaneously to each polymer by PYSIMM. Each polymer chain is then replicated six times and put in a simulation box with periodic boundary conditions. Subsequently, the structures will be optimized using the large-scale atomic-molecular massively parallel simulator (LAMMPS)[71]. The initialization process involves simulating the system under the NVT ensemble at 100 K for 2 ps, followed by a gradual heating up to 1000 K in 1 ns, and then simulated in the NPT ensemble for an additional 50 ps at 0.1 atm, achieving further relax the structure and eliminate close contacts between atoms. Following this, the system experiences a 1 ns NPT simulation at 1000 K, allowing the pressure to rise from 0.1 atm to 500 atm. Subsequently, the obtained polymer system was annealed from 1000 K to 300 K with a cooling rate of 140 K/ns in an NPT ensemble at 1 atm, followed by another NPT run at 300 K and 1 atm for 8 ns as the relaxation process to achieve the final equilibration state. This procedure is designed to simulate realistic polymer behaviors under diverse conditions.

**TC Calculation.** Each relaxed amorphous polymer structure, which was in a cubic box, is then duplicated in three copies to form a cuboid for TC calculation. Their sizes are around 9.9 × 3.3 × 3.3 nm³, with slight variations due to density differences of specific polymers. The cuboid is then employed for calculating TC via Non-Equilibrium MD (NEMD) simulations. The system is run in an NVE ensemble for 5 ns with a 0.25 fs timestep to capture the vibrational dynamics of light hydrogen atoms. In NEMD, we establish thermal gradients using Langevin thermostats at opposite ends of the system, setting a heat source at 320 K and a sink at 280 K. Finally, the heat flux and temperature profiles over the last 5 ns of the simulation are recorded to calculate TC using Fourier's law[72], $\kappa = -\frac{J}{\nabla T}$, where $J$ is the heat flux and $\nabla T$ is the temperature gradient along the heat flux direction.

**Modulus Calculation.** The bulk modulus of amorphous polymers was computed using a finite-deformation stress-strain method implemented in LAMMPS. Each equilibrated polymer cell was first converted from a cubic to a triclinic box to allow for shear deformation. Symmetric ±2% finite strains were then applied along the six independent strain modes (three normal and three shear) with atom remapping. After each ± deformation, the system was run at fixed volume at 300 K using NVE integration coupled with a Langevin thermostat, consisting of a 250-fs brief equilibration followed by a 75-fs sampling segment (0.25 fs timestep), and the time-averaged stress tensor was recorded. The elastic stiffness constants[73] $C_{ij}$ were derived from the linear relation between stress ($\sigma_i$) and strain ($\varepsilon_j$) within the elastic regime, yielding the complete 6 × 6 stiffness matrix via central finite differences. The bulk modulus was computed from the symmetrized stiffness constants using the Voigt expression $B = (C_{11} + 2C_{12})/3$, which is equivalent to the Voigt-Reuss-Hill value under isotropic symmetry. To ensure statistical reliability, for each simulation, the final mean bulk modulus was averaged from 100 snapshot evaluations evenly sampled over an 8 ns production trajectory. A representative polyacrylic acid example illustrating the snapshot-averaging stability is provided in Supplementary Fig. S2b. This workflow adheres to standard finite-temperature elasticity protocols in LAMMPS.

### 4.2 Polymer representation and DKL surrogate model

We constructed separate DKL surrogate models to predict TC ($k$) and bulk modulus ($B$). The DKL framework combines a neural feature extractor $\phi_\theta(\cdot)$ with a GP prior in the latent feature space:

$$\mathbf{z} = \phi_\theta(\mathbf{x}), \quad f(\mathbf{z}) \sim \mathcal{GP}(m(\mathbf{z}), k_\psi(\mathbf{z}, \mathbf{z}')), \quad y_i = f(\mathbf{z}_i) + \varepsilon_i, \quad \varepsilon_i \sim \mathcal{N}(0, \sigma_n^2)$$

where $\phi_\theta$ is an MLP-based FeatureNet and $k_\psi$ represents the GP kernel (Rational Quadratic kernel for TC and Matern kernel with $\nu = 1.5$ for modulus).

Input features derived from p-SMILES were converted into PE, the approximately 300-dimensional embeddings using mol2vec, and subsequently standardized based on the initial set for numerical stability. Target properties (TC in W m$^{-1}$ K$^{-1}$ and bulk modulus in GPa) were standardized separately (z-score space) to remove bias from differing scales. The feature extractor is an MLP with ReLU + dropout feeding an Exact GP: for TC - hidden (192, 128), dropout 0.1, latent 16; for Modulus - hidden (224, 192, 128, 32), dropout 0.05, latent 12. Model architectures were selected based on Optuna[74] tuning and then fixed for all subsequent experiments. The GP models used a ConstantMean function and ScaleKernel applied over an Automatic Relevance Determination (ARD) kernel matching the latent dimension. The TC surrogate utilized a Rational Quadratic kernel (with an alpha prior set as Gamma (1.5, 1.0)), while the modulus surrogate employed a Matern kernel with $\nu = 1.5$. Both GPs used a Gaussian likelihood function.

Training proceeded in two optimization phases under double-precision computation: an initial Adam pre-optimization phase (learning rate 3×10$^{-4}$ for 200 epochs) followed by L-BFGS fine-tuning (TC surrogate: learning rate 0.8 for 40 iterations; modulus surrogate: learning rate 0.8 for 50 iterations), with maximizing the exact marginal log likelihood in GPyTorch with joint end-to-end updates of the MLP weights/biases, GP kernel parameters, the constant mean, and the likelihood noise $\sigma_n^2$. We evaluated the models through 5-fold CV. All predictions were made using the observation-noise posterior, and uncertainties were back-transformed to physical units via their respective scalers. Reported CV metrics and parity plots are also shown in physical units. A consistent random seed (42) was applied throughout the entire process.

### 4.3 Multi-objective optimization and active learning

All multi-objective optimization was conducted in the z-score space used to train the DKL surrogates, with the bulk modulus sign-flipped ($-B$) so both objectives were maximized. We construct a two-objective surrogate by combining the independently trained DKL models for $k$ and $-B$ into a ModelListGP. Candidate selection is driven by the **$q$ -Noisy Expected Hypervolume Improvement ($q$NEHVI)** acquisition function, defined as

$$\alpha_{q\text{NEHVI}}(\mathbf{X}_q) = \mathbb{E}_{F_n, f(\mathbf{X}_q)}[\text{HV}(\text{ND}(F_n \cup f(\mathbf{X}_q))) - \text{HV}(\text{ND}(F_n))],$$

where $F_n$ denotes a posterior sample of the previously evaluated objectives, ND(·) the non-dominated set, $\mathbf{X}_q$ the batch of $q$ candidates, and HV(·) the hypervolume with respect to a reference point. Monte Carlo sampling approximation was used to estimate the joint posterior expectation for candidate selection (batch size $q = 4$) over the unlabeled polymer pool. Feature-wise min-max normalization bounds (10% padding) were computed from the labeled embeddings and used to normalize both the baseline set and the unlabeled pool for acquisition optimization, and the reference point was initialized 6 units ($\varepsilon$=6) below the initial objective minima and kept fixed for hypervolume (HV) computation (validated each iteration to remain below the current minima). Newly selected polymers were evaluated via MD to obtain ground-truth property labels, appended to the training set, and removed from the unlabeled pool to prevent duplication. Both DKL surrogates were retrained each iteration using the same configuration as in surrogate model construction. After every update, the updated HV was computed and 5-fold CV performed to monitor $R^2$ and MSE in real units. The AL-MOBO loop proceeded for 60 iterations, with the iteration achieving the highest HV designated as the final optimum ($S_{\text{best}}$).

We evaluated each MOBO iteration in terms of probabilistic performance using the Gaussian negative log-likelihood (NLL) and uncertainty calibration using the expected normalized calibration error (ENCE). At each iteration $t$, we prospectively evaluated performance on the newly acquired batch using surrogate models trained up to iteration $t-1$. Specifically, predictive means $\mu_i$ and standard deviations $\sigma_i$ were obtained from the posterior predictive distribution including observation noise. The NLL was computed as

$$\text{NLL} = \frac{1}{N}\sum_{i=1}^{N}\left(\frac{1}{2}\log(2\pi\sigma_i^2) + \frac{(y_i-\mu_i)^2}{2\sigma_i^2}\right),$$

where $N$ is the number of newly evaluated polymers at iteration $t$.

For calibration assessment, predictions were sorted by $\sigma$, partitioned into $B$ equal-sized bins (per iteration $B=4$, matching $q=4$). In each bin $b$, we calculated

$$\text{RMSE}_b = \sqrt{\frac{1}{|b|}\sum_{i\in b}(y_i-\mu_i)^2}, \quad \text{RMS}_{\sigma,b} = \sqrt{\frac{1}{|b|}\sum_{i\in b}\sigma_i^2},$$

and defined ENCE as

$$\text{ENCE} = \frac{|b|}{N}\sum_{b=1}^{B}\frac{|\text{RMSE}_b - \text{RMS}_{\sigma,b}|}{\text{RMS}_{\sigma,b}},$$

To stabilize small-batch variation, a rolling pooled ENCE was computed by pooling samples over a sliding window of $W=5$ iterations prior to binning. As ENCE depends on the employed binning strategy, we report it as an internal calibration diagnostic rather than an absolute benchmark. Learning-curve trends for NLL were smoothed for visualization using an exponentially weighted moving average (EWMA, $\alpha=0.35$), while reported statistics were computed from raw per-iteration metrics. All uncertainty metrics were computed in the standardized z-score space used during optimization.

### 4.4 Post-optimization interpretability and analyses

**t-SNE visualization.** We visualized the polymer feature space using t-SNE to examine how the initial set, unlabeled pool, and MOBO-selected candidates were distributed in the polymer embedding space. PEs were standardized using a scaler fitted on the combined initial and unlabeled datasets, and the same scaler was applied to all subsets. Dimensionality was reduced with PCA ($\leq 50$ components) before computing a single unified 2D t-SNE embedding. Perplexity was set adaptively based on the total sample size. All subsets were embedded jointly to ensure comparability and prevent drift between maps. Following standard practice, the resulting t-SNE map is interpreted qualitatively to reveal local chemical neighborhoods rather than global distances.

**Feature importance analysis.** For structure-property interpretation, we trained tree-ensemble regressors on the MD-labeled polymer set (all selected candidates combined with the initial set, totaling 306 polymers) and computed SHAP values to quantify feature importance. Input features were built from a hierarchical descriptor set using PolyMetriX at full-polymer, backbone, and sidechain levels, and were minimally supplemented with RDKit physicochemical descriptors computed at the same hierarchy, span size and shape, π-conjugation and rigidity, flexibility, polarity and hydrogen-bonding, polarizability, composition, and packing-related proxies. All descriptors were standardized (z-score). Highly correlated features were removed using a correlation threshold of $|r| > 0.92$, yielding a non-redundant subset of 82 features used for downstream analysis. For each target property ($k$ and $B$), we performed a small grid search over RandomForestRegressor and ExtraTreesRegressor hyperparameters and selected the best-performing model via 10-fold CV. Out-of-fold SHAP values were then computed with TreeExplainer (interventional mode) to quantify each descriptor's contribution to TC and bulk modulus. SHAP values are reported using a consistent sign convention: positive values indicate higher model-predicted $k$ or $B$. The interpretations are visualized through top K feature bar charts (K=15), value-colored SHAP beeswarm plots, and SHAP-feature dependence plots for key descriptors. For interpretability, Fig. 7a and 7b focuses on backbone/sidechain descriptors that directly map attributions to structural hierarchies. Descriptor stability was further assessed via bootstrap resampling (500 iterations) and y-scramble controls (20 permutations), by monitoring the overlap and Jaccard similarity of the Top-15 descriptor sets across resamples.

**Acknowledgements**

This work was supported in part by the Notre Dame Center for Research Computing and National Science Foundation grants (2332270 and 2102592).

**Author contributions**

Y.L. and T.L. conceived the project. Y.L. developed the methodology, implemented the model, and carried out the investigation. Y.L., J.X. and R.Z. curated the data. Y.L., T.L. and M.J. performed the formal analysis. Y.L. prepared the visualizations and wrote the original draft. T.L. supervised the work and acquired funding. All authors reviewed and edited the manuscript and approved the final version.

**Data availability**

The authors declare that the data supporting the findings of this study are available within the article and its supplementary information files or will be available for download from https://github.com/yuhanliu57/Polymer-ALMOBO upon publication.

**Code availability**

The code for this study will be available for download from https://github.com/yuhanliu57/Polymer-ALMOBO upon publication. Other codes can be available upon reasonable request from the authors.

**Conflict of interest**

The authors declare no competing financial interest.

**Additional information**

**Supplementary information** is available for this paper.